\documentclass[fleqn,usenatbib]{mnras}
\usepackage{graphics,epsf}
\usepackage{amsmath}                % American Mathematical Society package
\usepackage{amsfonts}               % American Mathematical Society fonts
\usepackage{amssymb}                % American Mathematical Society symbol
\usepackage{epsfig}                 % EPS figures
\usepackage{graphicx}
\usepackage{float}

\newcommand{\cm}{{~\rm cm}}
\newcommand{\km}{{~\rm km}}
\newcommand{\s}{{~\rm s}}

\newcommand{\g}{{~\rm g}}

\newcommand{\K}{{~\rm K}}

\newcommand{\yr}{{~\rm yr}}

\title[Variable jets]{Variable jets at the termination of the common envelope evolution}

\author[N. Soker]{
Noam Soker$^{1,2}$\thanks{E-mail: \href{mailto:soker@physics.technion.ac.il}{soker@physics.technion.ac.il}}
\\
$^{1}$Department of Physics, Technion, Haifa 3200003, Israel\\
$^{2}$Guangdong Technion Israel Institute of Technology, Shantou, Guangdong Province 515069, China\\
}

\begin{document}
\label{firstpage}
\pagerange{\pageref{firstpage}--\pageref{lastpage}}
\maketitle

\begin{abstract}
I propose that at the termination of the common envelope evolution (CEE) the companion to the giant star might launch jets that have variable directions and intensities, hence the jets shape the inner zones of the descendant nebula causing it to lose any type of symmetry.
This might account for some of the chaotic departures from axisymmetric structures in some planetary nebulae.
I base my study on the assumption that at the termination of the CEE the binary interaction forms a circumbinary thick disk that feeds the companion via an accretion disk that launches opposite jets. I consider two processes that might lead to the development of Rayleigh-Taylor instabilities that lead to chaotic accretion flow on to the companion. The chaotic accretion flow leads to a variable accretion disk around the companion, hence to variable jets.
In the first process the jets inflate low-density hot bubbles in the inner regions of the envelope. In the second process the inner boundary of the thick circumbinary disk, that faces the center, cools by radiation thus leading to a temperature, and hence also a pressure, increase with radius. The opposite pressure and density gradients lead to Rayleigh-Taylor instabilities. 
This study further underlines the importance of exploring the termination phase of the CEE to the understanding of the entire CEE and its outcomes. 
\end{abstract}


% ==========================================================
\section{INTRODUCTION}
\label{sec:intro}
% ==========================================================

The more than three decades-old idea that jets shape many planetary nebulae (PNe) (e.g., \citealt{Morris1987, Soker1990AJ, Soker1992, SahaiTrauger1998, LeeSahai2004}) had become more popular in recent years (e.g., \citealt{Dennisetal2009, Hugginsetal2009, Leeetal2009, Boffinetal2012, HuarteEspinosaetal2012, Balicketal2013, Miszalskietal2013, BlackmanLucchini2014, Tocknelletal2014, Huangetal2016, Sahaietal2016, RechyGarciaetal2016, GarciaSeguraetal2016, Balicketal2017, MorenoMendezetal2017, Akashietal2018, AkashiSoker2018, Chamandyetal2018, Franketal2018, ShiberSoker2018, LopezCamaraetal2019}). Since the formation of jets in evolved stars requires strong binary interaction, it is not surprising that during these more than three decades the community has reached the understanding that the asymptotic giant branch (AGB) progenitors of all non-spherical PNe are in close binary systems  (e.g, \citealt{Bondetal1978, BondLivio1990, SokerHarpaz1992, NordhausBlackman2006, GarciaSeguraetal2014, DeMarco2015, Zijlstra2015, DeMarcoIzzard2017}), including planetary systems (e.g., \citealt{SabachSoker2018} for a recent paper). 
 
The many studies of PNe with binary central stars have brought these systems to be the key to the understanding of the common envelope evolution (CEE), of the formation of other type of close compact binary systems, and of the formation of other types of bipolar nebulae. This community tour de force was made possible by many high quality observations of central binary systems of PNe, some that show jets, from which I list some papers only from the last four years (e.g., \citealt{Akrasetal2015, Alleretal2015a, Boffin2015, Corradietal2015, Decinetal2015, DeMarcoetal2015, Douchinetal2015, Fangetal2015, Gorlovaetal2015, Jonesetal2015, Manicketal2015, Martinezetal2015, Miszalskietal2015, Mocniketal2015, Montezetal2015, Akrasetal2016, Alietal2016, Bondetal2016, Chenetal2016, Chiotellisetal2016, GarciaRojasetal2016, Hillwigetal2016a, Jones2016, Jonesetal2016, Madappattetal2016, Chenetal2017, Hillwigetal2017,JonesBoffin2017, Sahaietal2017, Sowickaetal2017, Dopitaetal2018, Jonesetal2018, Miszalskietal2018}). These and other studies open more questions. 

One question arises from the inspections of the morphologies of PNe with close central binary stars (for an updated list see \citealt{Jones2018}). 
I find that many show chaotic perturbations to their axisymmetrical morphologies (e.g., M3-1, Hen~2-11, K~1-2, NGC~6026, Hen~2-155, Hen~2-161, NGC~6326). 
I see these departures from axisymmetry also in PNe where the orbital motion axis and the axis of the PN were found to be aligned, e.g., NGC 6778, and Abell 41 \citep{Hillwigetal2016b}.
In NGC~6778, for example, the jets posses a clear departure from axisymmetry.
From these two lists of PNe I estimate that about 10-15 per cent of PNe with post-CEE binary central stars show chaotic perturbations. From these PNe, the fraction is larger, about 20-25 per cent, for PNe with a clear equatorial ring.  
These chaotic features suggest that in many cases the launching process of jets in binary progenitor systems contain a chaotic component, e.g., the mass feeding of the accretion disk around the companion is chaotic and/or unstable. 

In the present paper I raise the possibility that at the termination of the CEE, or shortly after, accretion has a chaotic component that leads to variable jets.  Observations and their analysis show that in some cases jets are launched after the ejection of the main nebula, i.e., in the post-AGB phase (e.g., \citealt{Soker1992}).
\cite{Huggins2007} studied nine PNe and found that jets typically appear slightly later than the tori that the PN progenitors eject. For binary systems that experience CEE this implies that the jets are post-CE jets, as in NGC~6778 \citep{Tocknelletal2014}. 
Another possibility, due to \cite{Nordhausetal2007}, for late jets is that a companion spins-up the common envelope such that a dynamo activity in the envelope leads to the launching of jets at about 100 years after the main ejection of the envelope.

In this study I target binary systems that launch jets as they exit the CEE or shortly thereafter, and examine the processes by which the post-CEE jets acquire stochastic intensity and/or direction. 

In section \ref{sec:flow} I discuss the launching of jets in the post-CEE phase. In section \ref{sec:bubbles} I discuss the instabilities resulting from the inflation of hot bubbles by jets, while in section \ref{sec:temperature} I discuss the effect of an inverse temperature gradient. 
I summarise in section \ref{sec:summary}, where I conclude that numerical simulations that study the exit from the CEE must include energy transfer by radiation and convection. In a recent paper \cite{WilsonNordhaus2019} claim that it is important to include energy transfer, in particular by convection, when simulating the entire CEE. 
   
% ==========================================================
\section{JETS AT THE FINAL CEE PHASE}
\label{sec:flow}
% ==========================================================

At the late CEE phase the secondary star is deep in the envelope and most of the leftover envelope resides outside its orbit. The envelope possesses a highly oblate structure that might resembles more a torus like structure, i.e., a circumbinary disk, as \cite{KashiSoker2011MN} proposed and \cite{Kuruwitaetal2016} demonstrated in their hydrodynamical numerical simulation. \cite{KashiSoker2011MN} suggested that the binary system (the core and secondary star) gravitationally interacts with the circumbinary disk and as a result of that further spirals-in. \cite{ChenPodsiadlowski2017} followed \cite{KashiSoker2011MN} and further developed this mechanism. 
The inner boundary of this circumbinary disk can be steep, leaving an almost empty funnel along the polar directions, as numerical simulations of the CEE show (e.g., \citealt{Franketal2018, Reichardtetal2018}).
\cite{Kuruwitaetal2016} find that some mass is falling in from all directions, implying that the polar funnels are not completely empty. 
 
The relevant process to this study is that mass from the circumbinary disk flows toward the center. The secondary can accrete mass from this flow, most likely thought an accretion disk. Note that there are two disks in this scenario. A circumbinary disk and an accretion disk around the companion. The circumbinary disk is much larger than the accretion disk and feeds it with mass. Such an accretion disk launches jets. An interesting process of accreting mass from the circumbinary disk is that in some cases the angular momentum direction of the accretion disk around the secondary star can be opposite to the orbital angular momentum \citep{Soker2017}. 

This circumbinary disk is bound, i.e., it forms a semi-Keplerian disk (pressure within the disk can play a role), and the interaction can last for a time of weeks to months \citep{KashiSoker2011MN}. As the orbital period of the core-companion system is hours, the interaction lasts for a time of tens to hundreds of orbital periods. But the intensity of the interaction, like the accretion rate from the circumbinary disk to the binary system, can vary a lot.
Namely, the accretion process might be discontinuous such that there are many mass accretion episodes. 
 
There are two key processes that I examine. One is the role of the jets that the secondary star launches (section \ref{sec:bubbles}), and the second is radiative cooling of the circumbinary disk (section \ref{sec:temperature}). Numerical simulations do not include these processes because they are computationally demanding. 

\begin{figure}
\begin{center}
%\vspace*{-1.2cm}
%\hspace*{-4.2cm}
\includegraphics[trim= 7.8cm 16.5cm 0.0cm 4.0cm,clip=true,width=0.65\textwidth]{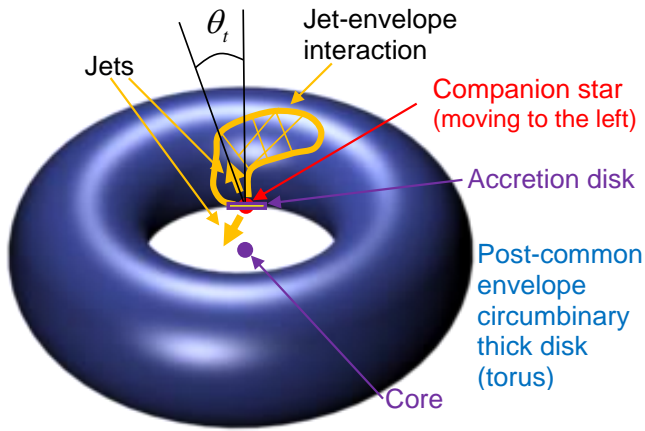}
%%%%  trim={<left> <lower> <right> <upper>}
%%% \includegraphics[scale=0.40]{PostCEGrazingFig1.pdf} \\
\vspace*{-0.5cm}
\caption{A schematic drawing (not to scale) of the last phase of the CEE, when the envelope might reside in a circumbinary disk that might last to the post-CEE phase. The secondary star accretes mass from the envelope outside its orbit and launches jets.  At the momentarily location of the secondary star in the figure it moves to the left and it launches jets perpendicular to the orbital plane in its frame. Because of the orbital motion the jets are tilted with respect to the center of mass in the direction of motion of the secondary star (equation \ref{eq:thetat}). The orange hatched region depicts the hot bubble formed by the interaction of one jet with the envelope (I do not show the opposite bubble). There is also lower-density gas in the entire volume, supplied also by in-falling gas, but it is not shown. }
\label{fig:cartoon}
\end{center}
\end{figure}

But I start by describing another effect that breaks axisymmetry and that I studied in an earlier paper where there are more details \citep{Soker2017}. 
The main sequence secondary launches jets with a velocity of about the escape velocity from its surface $v_j \simeq 500 \km \s^{-1}$, while at the same time it orbits the center of mass of the secondary-core binary system at a velocity of $v_{\rm orb}$. As a result of that the initial velocities of the two opposite jets have a component in the direction of this orbital motion. The angle of deflection relative to the orbital angular momentum axis is $\tan \theta_t = v_{\rm orb}/v_j$, as I draw schematically in Fig. \ref{fig:cartoon}.
Scaling the core mass with $M_{1c}=1M_\odot$ and the secondary mass with $M_2=0.3 M_\odot$, and taking a circular orbit with an orbital separation of $a$ yields  
\begin{eqnarray}
\tan \theta_t = \frac {v_{\rm orb}}{v_j} =
0.34
\left( \frac {v_j}{500 \km \s^{-1}} \right)^{-1}
\left( \frac {M_{1c}}{1 M_\odot} \right) \nonumber \\
\times
\left( \frac {M_{1c}+M_2}{1.3 M_\odot} \right)^{-1/2}
\left( \frac {a}{5 R_\odot} \right)^{-1/2} ,
  \label{eq:thetat}
\end{eqnarray}
corresponding to $\theta_t = 19^\circ$.
% FFFFFFFFFFFFFFFFFFFFFFFFFFFFFFFFFFFFFFFFFFFFFFFFFFFFFFFFFFFFFF
% FFFFFFFFFFFFFFFFFFFFFFFFFFFFFFFFFFFFFFFFFFFFFFFFFFFFFFFFFFFFFF

The initial tilt of the leading part of the jet is larger even because of the finite width of the jet (or disk wind). 
If, for example, the tilt of the jet's axis is $\theta_t=20^\circ$ and the half-opening angle of the jets is $\alpha_j=30^\circ$, then the leading edge of the jets, the edge toward the orbital direction, is at $50^\circ$ to the orbital angular momentum axis. 
Since the circumbinary disk is thick, it is very likely that at least the leading part of the jets, toward the orbital motion, interact with the inner boundary of the circumbinary disk. This can create relatively hot bubbles. The interaction of the jets with circumbinary disk causes another tilt, in a different direction, as in the interaction of jets with the outer envelope in the grazing envelope evolution
\citep{Shiberetal2017}. 

If an accretion episode lasts for a time that is shorter, or not much longer, than the orbital period, then the jets introduce disturbance to the outflowing gas that departs from axisymmetry. The effect of the orbital motion by itself still possesses a mirror symmetry with respect to the orbital plane. 

% ==========================================================
\section{Jet-inflated hot bubbles}
\label{sec:bubbles}
% ==========================================================

Consider the CEE stage when the companion is deep in the envelope, approaching its final orbital separation, and the density of the inner region is low due to rotation, mass loss, and the inflation of the envelope by the in-spiralling companion. The in-spiralling companion deposits angular momentum to the envelope, and the rotating envelope acquires an oblate shape. 
Two processes might act to turn the oblate envelope to a thick circumbinary disk, or a torus (although the shape does not need to be a mathematical torus). 
The first process is further deposition of angular momentum to the envelope, in particular to the inner regions of the envelope.  
The second one is the inflation of low-density high-temperature bubbles by jets that the companion might launch.  I now consider this process. 

Consider a main sequence companion that launches jets at the escape speed from the star, $v_j \simeq v_{\rm esc,2} \simeq 500 \km \s^{-1}$.
The jets are shocked to a temperature of 
\begin{equation}
T_{\rm bubble} = 
3.5 \times 10^6 
\left( \frac{v_{\rm jet}}{500 \km \s^{-1}} \right)^2  \K,
\label{eq:Tbuble}
\end{equation}
where I assume high mach number jets and an adiabatic index of $\gamma=5/3$ inside the jets, as appropriate for fully ionised gas.
The temperature of an AGB star in the zone of $r \simeq 1-10 R_\odot$ can be approximated by  
\begin{equation}
T_\ast \simeq 10^{6} 
\left( \frac{r}{3R_\odot} \right)^{-1} \K. 
\label{eq:Tstar}
\end{equation}
 
The temperature ratio of $T_{\rm bubble}/T_\ast$ implies that to maintain a pressure equilibrium the density ratio between the bubbles and stellar environment is $\rho_{\rm bubble}/\rho_\ast \simeq (T_{\rm bubble}/T_\ast)^{-1}$, and hence the mass in the bubbles of volume $V_{\rm bubble}$ is $M_{{\rm bubble},V} \simeq (T_{\rm bubble}/T_\ast)^{-1}  M_{\ast,V}$, where $M_{\ast,V}$ is mass of the undisturbed stellar envelope inside the same volume. If the jets carry a fraction of  $\epsilon_j$ of the accreted mass, then to inflate the bubbles the companion should have accreted a mass from a stellar volume $V_{\rm accrete}$ that is given by
 \begin{equation}
\frac{V_{\rm accrete}}{V_{\rm bubble}} \simeq 3 
\left(\frac{\epsilon_j}{0.1} \right)^{-1}
\left( \frac{r}{3R_\odot} \right)^{-1} 
\left( \frac{v_{\rm jet}}{500 \km \s^{-1}} \right)^{-2} .
\label{eq:Volume}
 \end{equation}
This is quite possible at $ r \ga 10 R_\odot$ for the parameters I use here, as the volume of the accreted gas is about equals to the volume of the jet-inflated bubbles. Furthermore, as the envelope evolves toward a torus-like structure the density along the polar directions where the jets inflate the bubbles becomes smaller. Hence, the required mass in the bubbles becomes smaller, and so is the demand for the accreted mass. This implies that the inflation of bubbles can be important down to an orbital separation of several solar radii.  

I consider accretion and the inflation of bubbles only at the termination of the CEE, when the densities in the inner regions are small, so the companion can accrete the required mass at a not too high rate. Namely, I assume that the main sequence companion does not expand much as it accretes mass from the envelope.  It accretes a mass of $M_{\rm acc} < 0.1 M_\odot$ and launches jets with a mass of $M_{\rm jets} \approx 10^{-5}-0.01 M_\odot$. Even though they have low mass, due to their  velocity which is tens times the nebular velocity, the jets might influence the slow wind from the AGB star. 
 
Consider then the situation where the companion launches jets as it orbits the core and that the jets inflate bubbles. The low-density hot bubbles displace the stellar gas. The situation where a low density gas supports a higher density gas is prone to the Rayleigh-Taylor instability, that its growth time-scale (the e-folding time) can be written as  
\begin{equation}
t_{\rm RT} \simeq (2 \pi )^{-3/2} 
\left( \frac{\lambda}{r} \right)^{1/2}
\left( \frac{\rho_\ast-\rho_{\rm bubble}}{\rho_\ast + \rho_{\rm bubble}} \right)^{-1/2}
t_{\rm Kep}  ,
\label{eq:tRT}
 \end{equation}
where $t_{\rm Kep}$ is one orbital period and $\lambda$ is the wavelength of the perturbation. For $\lambda \approx r$, this gives that in one orbit the instability growths by a factor of $\approx e^{10}$. 
At the same time the bubbles buoy up at a speed that is a fraction of the sound speed. In one orbit the bubbles can buoy up a distance of about the orbital separation.  
Overall, the growth of the instability implies that when the companion completes one orbit and returns to the place where it inflated bubbles before, the medium from where it accretes mass might be very clumpy due to the instabilities. The accretion process, and hence the accretion disk, might be variable, both in intensity and in the direction of the angular momentum axis. If the accretion disk launches jets, they also have variable direction around the orbital angular momentum and have a variable intensity.        
  
To leave an imprint on the descendant nebula the jets must break out from the inner envelope. The idea is that as time progresses the polar directions are cleared and the jets are free to interact with the slower wind. The chaotically variable jets leave chaotic morphological features in the inner regions of the descendant nebula.
Later ionisation by the central star can alter the structure, but cannot completely remove the chaotic signatures. 

This very complicate flow structure near the core-companion binary system is more complicated even because of the rotation of the envelope and the very likely situation where the envelope is convective. 
Only three-dimensional hydrodynamical numerical simulations can give us the answer whether the somewhat speculative mechanism I propose here leads to variable jets, and whether during part of the time these jets break out from the inner envelope regions so as to leave an imprint on the descendant nebula. 
 
% ==========================================================
\section{Inverse temperature gradient in the circumbinary disk/torus}
\label{sec:temperature}
% ==========================================================

Consider an ideal structure of a circumbinary torus (but not a mathematically perfect torus), or a thick accretion disk. The companion just removed the inner part of the envelope and pushed it further out. The inner edge of the torus becomes a more or less free surface that faces the center and starts radiating approximately as a black body. The central star heats the inner surface. If the inner surface of the torus faces the central star and fully absorbs the central star radiation, then at equilibrium the temperature of the inner surface would be 
\begin{equation}
T^4_{\rm i,eq} \le \frac{L_\ast}{ 4 \pi \sigma r^2_{\rm i}} ,
\label{eq:Tieq}
\end{equation}
where $L_\ast$ is the luminosity of the central star, $\sigma$ is the Stefan–Boltzmann constant, and $r_{\rm i}$ is the radial distance of the inner boundary from the star. 
The value of $r_{\rm i}$ increases as the segment of the inner boundary is further from the equatorial plane. As the surface might be inclined, the temperature is lower even than what the equality sign gives, hence the inequality sign. 
 
During the giant phase, before the companion disturbed that region, the temperature in that radius is given by energy transfer. Since most energy is carried by convection, the temperature is somewhat lower than that given by radiative transfer, but not by much 
 \begin{equation}
T^4_{\rm i,\ast} \la \frac{3}{16} \frac{l_T}{l_{\rm mean}} \frac{L_\ast}{ 4 \pi \sigma r^2_{\rm i}} \gg T^4_{\rm i,eq},
\label{eq:Tiast}
\end{equation}
where $l_T \equiv T (dT/dr)^{-1}$ is the temperature scale height and $l_{\rm mean} = (\kappa \rho)^{-1}$, where $\kappa$ is the opacity and $\rho$ the density, is the photon mean free path. The last inequality comes from the relation $l_{\rm mean} \ll l_T$.   
Examining some AGB models show that for $r_{\rm i} \simeq {\rm several} \times R_\odot$ the ratio is $T_{\rm i,\ast} \approx 10 T_{\rm i,eq}$. 
  
After the binary interaction forms the circumbinary disk/torus, the inner surface of the torus starts to cool. As a result of that an inverse temperature gradient develops, i.e., opposite to that in the star, and the magnitude of the pressure gradient decreases, and gravity overcomes pressure gradient. In other words, the zones further away from the inner boundary of the torus pushes the inner regions inward. Mass from the inner regions can then be accreted.  

Consider typical numbers for a low mass companion orbiting the core of an AGB star at an orbital separation of $a \approx 4 R_\odot$, i.e., an orbital period of $P_{\rm orb} \approx 1~{\rm day}$.
The density in that region is $\approx 10^{-6} - 10^{-5} \g \cm^{-1}$. For an opacity of $\kappa \approx 1 \cm^2 \g^{-1}$ the photon diffusion time over a distance equals to $a$ is 
$t_{\rm diff} \approx \kappa \rho a^2/c \approx 1~{\rm month} - 1 \yr$. During these few months some of the inner regions of the torus are pushed inward as they cools, and the companion accretes some of the mass in these regions. 

The density profile of an AGB star in that region is $\rho \propto r^{-\beta}$ where $\beta \simeq 2 -3 $. This implies that the density scale height is $l_\rho \simeq 0.4r$. Over a radial distance from $r_i$ to $1.5 r_i$ the density decreases by about a factor of $\rho(r_i) \simeq 3 \rho(1.5r_i)$. At early times, before the zones further from the inner boundary of the torus had time to cool, the reverse temperature gradient might be larger than the density gradient,i.e., $T(r_i) < T(1.5r_i)/3$. This is based on the ratio  $T_{\rm i,\ast} \approx 10 T_{\rm i,eq}$. 
This implies that the pressure gradient and the density gradient have opposite signד, in at least some zones behind the inner boundary of the circumbinary torus and in at least part of the time. The opposite pressure and density gradients make these zones prone to the Rayleigh–Taylor instability. As the centrifugal force mostly acts to stabilise such zones, not all zones with opposite density and pressure gradient will develop the instability sufficiently fast. I do expect some regions to become unstable and rapidly evolve to the non-linear regime. Only three dimensional simulations can determine the evolution of the instabilities. As with the Rayleigh–Taylor instabilities, tongues of dense gas are pushed into the low density gas, and low density tongues into the higher density regions. The instability might form dense clumps that are pushed toward the center and later are accreted on to the companion. 

The back-flow process that results from an inverse temperature gradient operates not only near the equatorial plane, but also away from the equatorial plane. The Rayleigh-Taylor instability that is likely to accompany this process leads to stochastic accretion that at a given time is not necessarily equal on the two sides of the equatorial plane. Namely, at a given time more mass might stream on to the companion that launches the jets from one side of the equatorial plane than from the other. This flow breaks the mirror symmetry about the equatorial plane such that if an accretion disk forms it is tilted to the equatorial lane. The tilted accretion disk launches jets with a symmetry axis that is inclined to the orbital angular momentum axis. Both the intensity and direction of the jets might change in a stochastic manner. More than that, the inflow from zones above the equatorial plane might disturb the jet on one side more than on the other side, further increasing the stochastic variation of the outflow between the two sides of the equatorial plane. 

% ==========================================================
\section{SUMMARY}
\label{sec:summary}
% ==========================================================

I discussed some jet-driven post-CEE processes that might cause a chaotic departure from the axisymmetric structure in the inner zones of post-CEE nebulae, such as observed in some PNe (section \ref{sec:intro}). 
I based my arguments for these processes on the assumption that at the termination of the CEE the binary interaction forms a circumbinary thick disk (or a torus) that feeds the companion via an accretion disk that launches opposite jets \citep{KashiSoker2011MN}.  
I summarise the processes as follows. 

\textit{Orbital motion} (section \ref{sec:flow}). The secondary star orbits the center of mass as it accretes mass and launches jets from the circumbinary torus (Fig. \ref{fig:cartoon}). If an accretion episode lasts for about an orbital period or less, the respective jets-launching episode does not possess an axisymmetry. This might repeat itself several times, adding a stochastic morphological feature. This process on its own does maintain a mirror symmetry about the equatorial plane, assuming the two jets are equal. 

\textit{Jet-inflated bubbles} (section \ref{sec:bubbles}). This process takes place if the jets inflate low-density hot bubbles not much smaller than the orbital separation. The low density bubbles displaced the cooler gas and support denser layers above them. This structure is prone to the Rayleigh-Taylor instability that can form a clumpy region. In the next orbit the companion might accrete gas from this clumpy region. The clumpy accretion process might form an accretion disk that launches jets with variable direction and intensity. The accretion disk might be inclined to the orbital plane, hence launching jets that break the mirror symmetry of the outflow.  
The situation is much more complicated due to the rotation of the envelope and likely convection in the envelope. Only three-dimensional hydrodynamical numerical simulations can reveal the true nature of this process, and whether indeed it might lead to variable jets.    

\textit{Inverse temperature gradient} (section \ref{sec:temperature}). When the density in the inner region, where the core-companion binary systems resides, becomes very low such that the region becomes optically thin, the inner boundary of the circumbinary thick disk cools by a more or less black body radiation. Although the core illuminates this surface, the temperature in the surface is much lower than the temperature of the gas at the same radius during the AGB phase. This implies that the temperature in the disk at larger radii is higher than on the boundary facing the center. The temperature increases with radius in that region close to the inner boundary of the circumbinary disk, inverse to the situation of a star. If this inverse temperature gradient becomes larger than the value of the density gradient then an inverse pressure gradient develops, pushing gas from near the inner boundary of the disk toward the center. Because the density and pressure gradient have opposite sense the region becomes Rayleigh-Taylor unstable. Some zones becomes unstable and can lead to accretion of dense clumps on to the companion in a stochastic manner. The result might be stochastic variation in intensity and direction of the jets. This mechanism breaks the mirror symmetry. The centrifugal force in the disk might stabilise some regions. 
This complicated process, or chain of processes, requires 3D hydrodynamical simulations that include energy transfer by radiation and convection, as \cite{WilsonNordhaus2019} claim should be for all stages of the CEE. 
   
In the three processes listed above the jets must break out from the inner envelope regions and interact with the slow wind to leave any morphological signatures in the descendant nebula. This in turn requires that some of the jets are launched at the very end of the CEE, after the polar directions have been cleared from most of the original envelope. 

This study further underlines the importance of the termination phase of the CEE to the general understanding of the entire CEE and its outcomes, e.g., final orbital separation and the morphology of the descendant nebula. For numerical simulations to follow these chaotic processes they will have to include energy transfer, both by convection and photons, and the luminosity from the central star. 

I thank Amit Kashi and an anonymous referee for helpful comments. 
This research was supported by the Israel Science Foundation.

\label{lastpage}
\end{document}